\documentclass[twocolumn,showpacs,amsmath,amssymb]{revtex4}
\usepackage{appendix}
\usepackage{xcolor}
\usepackage{graphicx}
\usepackage{dcolumn}
\usepackage{bm}
\def\b{\begin{equation}}
\def\e{\end{equation}}
\begin{document}
\title{Magnetostatic waves in metallic rectangular waveguides filled with uniaxial negative permeability media}

\author{Afshin Moradi$^{1}$}
\email{a.moradi@kut.ac.ir}
\author{Mohammed M. Bait-Suwailam$^{2,3}$}
\email{msuwailem@squ.edu.om}
\affiliation{%
\emph{$^{1}$Department of Engineering Physics, Kermanshah University of Technology, Kermanshah, Iran\\ $^{2}$Department of Electrical and Computer Engineering, Sultan Qaboos University, Muscat, Oman \\ $^{3}$Remote Sensing and GIS Research Center, Sultan Qaboos University, Muscat, Oman }
}%

\begin{abstract}
The propagation characteristics of magneto-quasistatic waves, more commonly, known as magnetostatic waves in a long, metallic rectangular waveguide filled with a metamaterial slab are comprehensively investigated. The metamaterial slab consists of split-ring resonators as an anisotropic uniaxial medium with transversal negative effective permeability. Some analytical relations and numerical validations on the characteristics of these waves are presented. The results include the dispersion relations, mode patterns (field distributions) that can be supported by such media and their corresponding cutoff frequencies, group velocities, power flows, and storage energies of magnetostatic waves. The findings from the present research study can be advantageous to advance the synthesis and development of negative permeability materials with peculiar features in guiding structures.
\end{abstract}

 \maketitle

\section{Introduction}
A metallic rectangular waveguide (as a basic guiding structure in microwaves, radars and antenna technology) is a long hollow tube (compared with its cross section) of rectangular cross section with four metallic walls. There are quite a number of interesting applications that the proposed structure can be useful and aid in the advancement of waveguides and antenna structures alike, including but not limited to miniaturization of waveguides \cite{S.H110, W.L2705}, performance enhancement of antennas integrated with anisotropic negative permeability media \cite{H.T1, Y.L737}, beamforming and filters using magnetostatic waves-enabled antennas \cite{A.R11}, and feeding structure for antenna arrays. 
\begin{figure}[!htb] 
\centering
\includegraphics[scale=0.31]{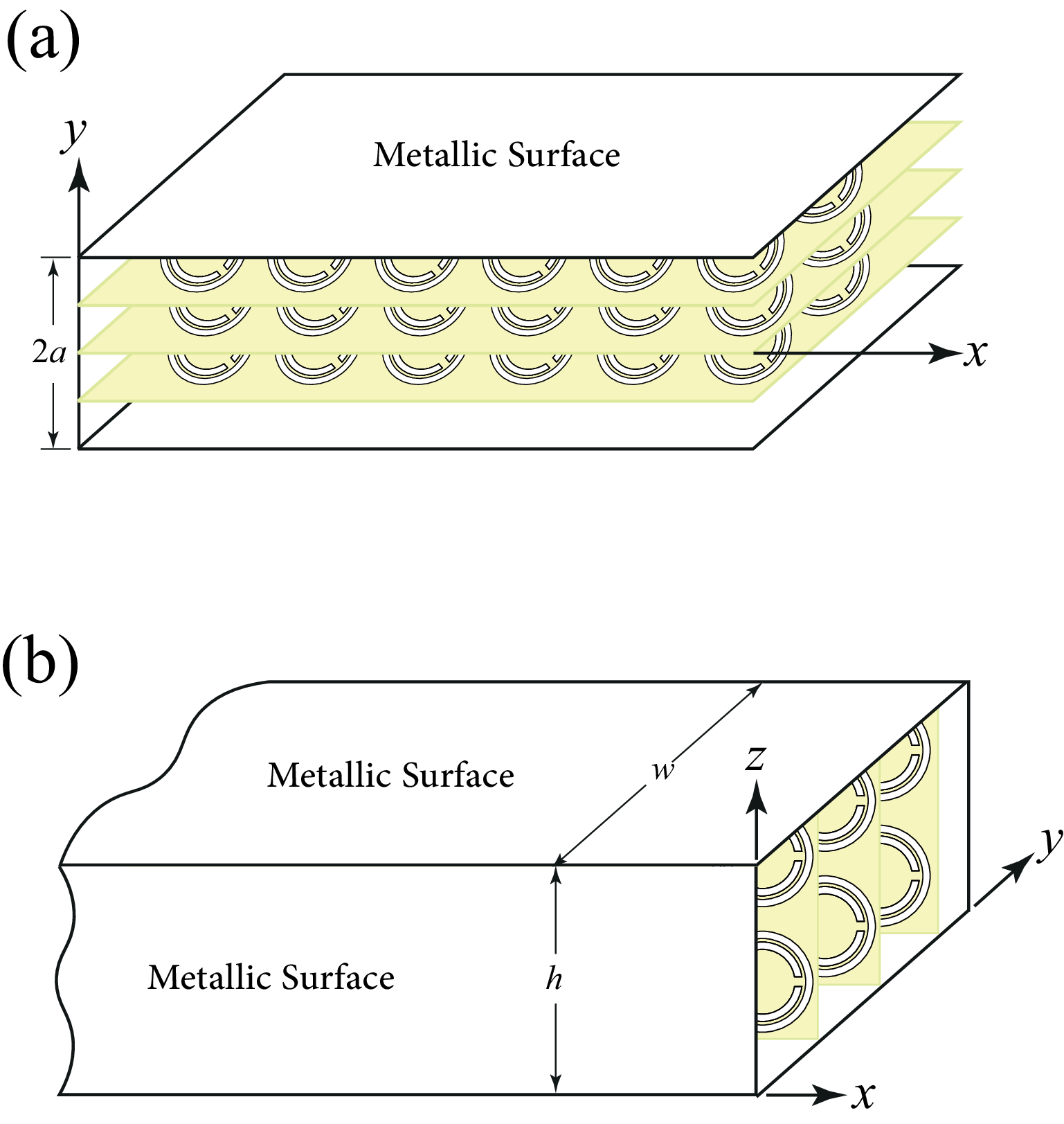}
\caption{(a) Side view of a planar metamaterial slab composed of SRRs as an anisotropic uniaxial $\mu$-negative
medium with two metallic boundaries. The metallic interfaces separating the slab ($-a\leq z\leq a$). (b) A metallic rectangular waveguide filled with a medium of SRRs with its appropriate dimensions.}
\label{fig.1} 
\end{figure}

Let us note that the investigation of electromagnetic characteristics of metallic waveguides loaded with various anisotropic media has been receiving considerable interests for almost seven decades \cite{A.T.M305, H.S409, G.B685, B.A.A4081, H.S.T134,S.L111, S.K.J346, K.S3429}. Also, the mode patterns (field lines) in hollow (rectangular/circular)  waveguides and cylindrical dielectric waveguides were presented in \cite{C.L271} and \cite{T.X2599}, respectively. Furthermore, there are many interesting works in (rectangular/circular) waveguides filled with metamaterials \cite{R.M183901, R.M405, Y.X426, S.H110, A.T2706, D.R2513, A.B3064, A.B3385, A.B677}. For instance, Marques \textit{et al.} \cite{R.M183901, R.M405} found a propagation below the cutoff frequency
of a very unusual waveguide loaded with so-called split-ring resonators (SRRs) that usually used for synthesis of metamaterial with negative permeability \cite{J.B.P2075, D.R.S4184, L.S, M.M.S2894}, and then extensively studied
by Hrabar \textit{et al.} in \cite{S.H110, S.H2587, S.H494}, where the results in \cite{S.H110} shows that backward propagation can occur when the longitudinal permeability is positive and the transversal permeability is negative. In fact, SRRs are an array of non-magnetic conducting rings that are arranged in a periodic fashion, which exhibit a strong resonant response on the magnetic component of electromagnetic field. As a result, these elements show an effective negative permeability \cite{A.B.S}.
\begin{figure}[!htb]
\centering
\includegraphics[width=8.5cm,clip]{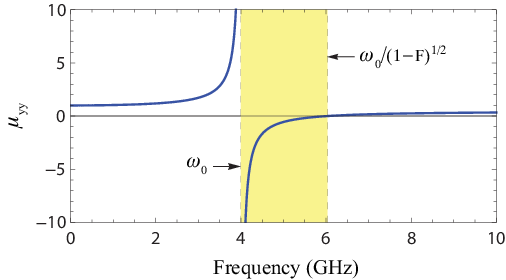}
\caption{Variation of $\mu_{yy}$ with respect to the frequency, when $\omega_{0}=4$GHz and $F=0.56$. The bandwidth of operation is
right above the resonance frequency $\omega_{0}=4$GHz to $\omega_{0}/\sqrt{1-F}\approx 6$GHz.}
\label{fig.2}
\end{figure}

Although the electromagnetic characteristics of metallic waveguides loaded by SRRs metamaterials have already been studied by several researchers \cite{S.H110, R.M183901, R.M2572,J.D.B1451, P.M36622, S.H2587, S.H494, M.W.F2955, F.Y.M3400}, to the best of our knowledge, the authors considered only the propagation of TE$_{m0}$ modes related to the waveguide axis, i.e., the $x$-axis in Fig. \ref{fig.1} (for instance, see \cite{S.H110, R.M183901, F.Y.M3400}). Furthermore, in the general case the axis of the waveguide does not coincide with the axis of TM and TE decomposition, and thus, a change of coordinates should be performed and analyzed. 

In earlier studies, the subject of electrostatic waves in metallic rectangular waveguides loaded by hyperbolic metamaterials and/or microwire metamaterials as anisotropic $\varepsilon$-negative media was extensively investigated and analyzed in \cite{A.M24522, A.M2024}. The analysis of magnetostatic waves on circular waveguides filled with anisotropic $\mu$-negative media were carried out in \cite{A.M143901, A.M178}.  In this study,  we investigate the propagation of magneto-quasistatic TE$_{y}$ waves (i.e., magneto-quasistatic TE waves related to the $y$-axis  in Fig. \ref{fig.1}) in a long, metallic rectangular waveguide filled with a metamaterial slab composed of SRRs as an anisotropic uniaxial $\mu$-negative medium. It is worth noting here that the characteristics of such magnetostatic waves were not tackled by neither earlier studies nor in recent works \cite{A.M143901, A.M178}.

Such magnetostatic waves do not exist
in the case of a hollow metallic waveguide or a metallic waveguide filled with an isotropic medium. Physically, the  magnetostatic waves (in general, quasi-static waves) owe their existence to the anisotropic property of the system and deserves particular attention.
In fact, in the absence of anisotropic $\mu$-negative properties, these slow magnetic waves disappear and cannot propagate.
\begin{figure}[!htb]
\centering
\includegraphics[width=7.5cm,clip]{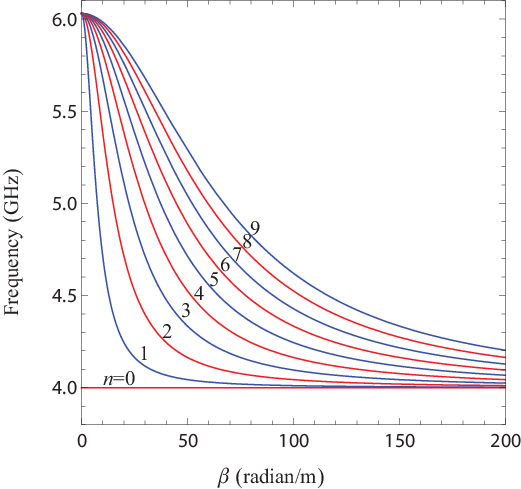}
\caption{Dispersion characteristics of the first ten bulk magnetostatic modes of a planar slab of SRRs with $a=25$mm, $\omega_{0}=4$GHz and $F=0.56$ and metallic boundaries, using Eq. \eqref{9}. The modes $n=0,2,4,6,8$ (red curves) are even (symmetric) modes and $n=1,3,5,7,9$ (blue curves) are odd (anti-symmetric) modes. The slope of curves is positive in the frequency
range $\omega_{0}<\omega<\omega_{0}/\sqrt{1-F}$ for all mode orders. Therefore, these modes called backward waves. For a backward wave, the directions of group velocity (power flow) and phase velocity (phase propagation) are mutually opposite. }
\label{fig.3}
\end{figure} 

Moreover, we note here that the presented study investigates thoroughly the dispersion characteristics and modal solutions of magnetostatic waves that exist in long waveguide structures filled with planar $\mu$-negative SRR media, which earlier studies, for example \cite{S.H110, A.M24522} did not investigate. In other words, our contribution in this research work is the presentation of a complete analytical and numerical solutions of magnetostatic waves in long metamateirals-based waveguide structures. In fact, the presented graphical solutions in this study adds more physical insights into the modal solution of magentostatic waves in engineered waveguide structures. Also, we should stress here that the important topics of energy density and power flow of magnetostatic waves in long waveguide structures filled with planar $\mu$-negative SRR media are the subjects of the last part of the present study.
 
\begin{figure}[!htb]
\centering
\includegraphics[width=9cm,clip]{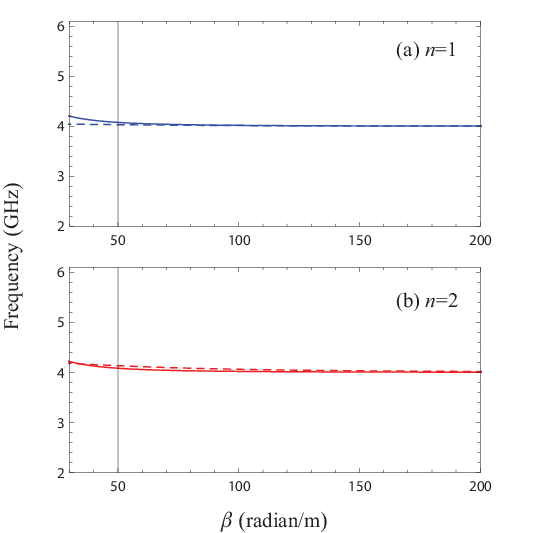}
\caption{Dispersion characteristics of the two modes ($n=1$ when $a=18.5$mm, and $n=2$ when $a=35.5$mm) of a planar slab of SRRs with $\omega_{0}=4$GHz and $F=0.56$ and metallic boundaries. Solid curves show the magnetostatic result using Eq. \eqref{9}. Dashed curves show the rigorous electromagnetic analyze.}
\label{fig.4}
\end{figure}

\section{Spectroscopy of magnetostatic modes of a planar slab of SRRs}
One type of magnetostatic waveguide is a planar slab of height $2a$ with two metallic boundaries filled with a medium of SRRs as an anisotropic uniaxial
metamaterial with transversal negative effective permeability, as shown in panel (a) of Fig. \ref{fig.1}. To simplify the analysis of the structure, at first we reduce the problem to a 2D one (its length in the $z$-direction is infinite) so that $\partial /\partial z=0$. We now investigate the behaviors of magnetostatic bulk modes of a planar slab of SRRs, based on the field analysis. Let us assume that the waves are traveling in the $x$-direction, and the structure is infinite in that direction as illustrated in panel (a) of Fig. \ref{fig.1}. We shall obtain the general expression for dispersion relation of the magnetostatic waves. For the present problem, the relative permeability tensor of the present medium is (see Eq. (6.1) in Chap. 6 of \cite{L.S})
\begin{equation}\label{1} \underline{\mu}(\omega)=\left(%
\begin{array}{ccc}
1 & 0 & 0\\
  0 & \mu_{yy} & 0\\
  0 & 0 & 1\\
\end{array}%
\right)\;,\end{equation}
where $\mu_{yy}$ is the relative permeability in the $y$-direction and has the form
\b \label{2} \mu_{yy}=1-\frac{F\omega^{2}}{\omega^{2}-\omega_{0}^{2}}\;,\e
in the lossless case \cite{A.M634}. Here $\omega_{0}$ is the resonance frequency of the SRRs and $F$ with $0 < F < 1$ is a measure of the strength of the interaction between
the SRRs and the magnetic field. One can see that Eq. \eqref{2} contains a
resonance frequency $\omega_{0}$ that allows $\mu_{yy}$ to
be negative over a bandwidth of nearly 2 GHz, depending on the applied field frequency $\omega$, as shown in Fig. \ref{fig.2}, when $\omega_{0}=4$GHz and $F=0.56$. 

Under the magnetostatic approximation with the magnetostatic equations $\nabla\times\textbf{H}=0$, $\nabla\cdot\textbf{B}=0$ and $\nabla\times\textbf{E}=-\partial \textbf{B}/\partial t$ \cite{D.D.S}, the magnetic field $\textbf{H}$ can be represented by the gradient of a magnetostatic potential $\psi$, as $\textbf{H}=-\nabla\psi$ \cite{A.M045801}. The Maxwell's equation $\nabla\cdot\textbf{B}=0$ with $\textbf{B}=\underline{\mu}\cdot\textbf{H}$ gives the wave equation for the magnetostatic potential of an unbounded medium of SRRs 
\b \label{3} \left[\dfrac{\partial^{2}}{\partial x^{2}} +\mu_{yy}\dfrac{\partial^{2}}{\partial y^{2}} \right]\psi=0 \;,\e
where all the field quantities are assumed to have
the harmonic time dependence of the form $\exp(j\omega t)$. Note that $\omega$ is angular frequency of a magnetostatic wave in the system.

In deriving the magnetostatic mode spectrum of a planar slab of SRRs, we use the following general separated-variable solution for traveling waves
in the $+x$-direction 
\b \label{4}
\psi(x,y)=\left( A \sin \kappa y+B \cos \kappa y\right)\exp(-j\beta x) \;.\e
Substituting this into Eq. \eqref{3}, we get  
\b \label{5} \beta^{2} +\mu_{yy}\kappa^{2}=0 \;.\e Eq. \eqref{5} shows that a homogeneous bulk magnetostatic plane wave corresponding to real $\beta$ and $\kappa$ is possible only when $\mu_{yy}$ is negative. Field existing within this waveguide must be characterized by zero normal components of magnetic field at the metallic walls. However, the boundary conditions at the two metallic walls can be written as 
\b \label{6} \dfrac{\partial \psi}{\partial y}\bigg\vert_{y=-a,a}=0\;. \e
Now, by applying the mentioned boundary conditions we can find the relationship
between the constants $A$ and $B$, as  
\b \label{7}
 A\cos\kappa a \pm B \sin\kappa a=0\;.\e 
To satisfy Eq. \eqref{7} the categories of even
and odd modes can be considered. For an even mode we have $A=0$ and then $ \kappa a=n\pi $
(with $n=0,1,2,\cdots$) is the dispersion relation for magnetostatic waves with symmetric potential functions. For an odd mode we have $B=0$ and therefore $\kappa a=\left( n+1/2\right)\pi $ is the dispersion relation for magnetostatic waves with anti-symmetric potential functions. The general dispersion relation can be written as 
 \b \label{8}\kappa a=n\dfrac{\pi}{2}  \;, \e
where this equation corresponds to an even mode for $n$ even, and to an odd mode for $n$ odd. Now, from Eqs. \eqref{2}, \eqref{5} and \eqref{8}, we obtain \begin{figure}[!htb]
\centering
\includegraphics[width=9.5cm,clip]{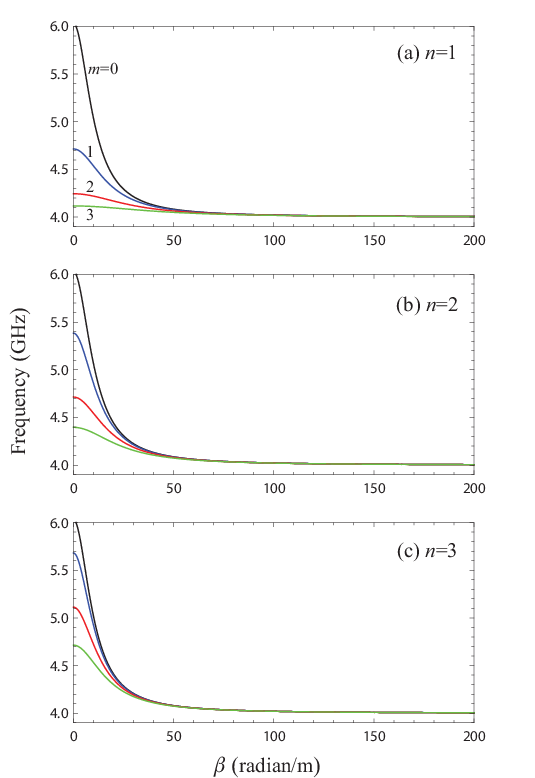}
\caption{Dispersion curves of magnetostatic modes ($m,n$) of a metallic square waveguide ($h=2a=w$)
filled with a medium of SRRs, using Eq. \eqref{17} for $\omega_{0}=4$GHz, $F=0.56$, $n=1,2,3$, and $m=0,1,2,3$. In panel (a) solid black, blue, red and green curves show the magnetostatic modes ($0,1$), ($1,1$), ($2,1$) and ($3,1$), respectively, when $a=18$mm. In panel (b) solid black, blue, red and green curves show the magnetostatic modes ($0,2$), ($1,2$), ($2,2$) and ($3,2$), respectively, when $a=35$mm. In panel (c) solid black, blue, red and green curves show the magnetostatic modes ($0,3$), ($1,3$), ($2,3$) and ($3,3$), respectively, when $a=55$mm. }
\label{fig.5}
\end{figure}  
\b \label{9} \omega=\omega_{0}\left(\dfrac{1+\left(\dfrac{n \pi }{2\beta a} \right) ^{2}}{1+(1-F)\left(\dfrac{n \pi }{2\beta a} \right) ^{2}}\right) ^{1/2}  \;.\e
 If $\beta\rightarrow0$, then $\omega\rightarrow\omega_{0}/\sqrt{1-F}$. If $\beta\rightarrow\infty$, then $\omega\rightarrow\omega_{0}$. This means that this system represents a backward bulk magnetostatic band filter with the angular frequency bands $1<\omega/\omega_{0}<1/\sqrt{1-F}$, as illustrated in Fig. \ref{fig.2}. Furthermore, all backward modes have equal cutoff frequency $(\omega_{c})_{n}=\omega_{0}/\sqrt{1-F}$, except $n=0$ mode. Actually, for $n=0$ the dispersion curve degenerates into a line and the wave stops propagating. 

Using Eq. \eqref{9}, the dispersion characteristics of magnetostatic modes of a planar slab of SRRs with $a=25$mm, $\omega_{0}=4$GHz and $F=0.56$ \cite{L.S}  for various values of $n$ are depicted in Fig. \ref{fig.2}. One can see an infinite number of bulk magnetostatic waves with different values of $n$ which depend on the conditions of excitation.  Also, we see that the frequency of a magnetostatic mode decreases monotonically with wavenumber throughout the allowed frequency range $\omega_{0}<\omega<\omega_{0}/\sqrt{1-F}$. Furthermore, in Fig. \ref{fig.4} the magnetostatic results shown in Fig. \ref{fig.2} is compared with the results using the rigorous
electromagnetic analyze for modes $n=1$ and $2$. In fact, by using the full set of Maxwell's equations, Eq. \eqref{5} for TE$_{0n}$ modes related to the $y$-axis should be read as (see Appendix)
 \b \label{5f} \beta^{2} +\mu_{yy}\left[ \kappa^{2}-\dfrac{\omega^{2}}{c^{2}}\right] =0 \;,\e
where $c$ is speed of light in free space. The comparison of the magnetostatic
results using Eq. \eqref{9}, with Eq. \eqref{5f} demonstrates
that for the present example the magnetostatic waves almost disappear for $\beta<50$ radian/m and cannot propagate. Physically, the existence of the magnetostatic waves (and also the electrostatic waves \cite{A.M24522}) is tied to the resonance frequency of the system, where the propagation constant (wavenumber $\beta$)  becomes infinity, here almost for $\beta>50$ radian/m. Near the resonance frequency, where the phase velocity $v_{\mathrm{ph}}$ [using Eq. \eqref{5f}] goes to zero, i.e.,
\b \label{5ff} v_{\mathrm{ph}}=\dfrac{\omega}{\beta}=\dfrac{c}{\beta\sqrt{\mu_{yy}}}\sqrt{\beta^{2} +\mu_{yy} \kappa^{2}}=0\;, \e
which leads to Eq. \eqref{5}, the magnetostatic waves can propagate in the system. In fact, the existence of the magnetostatic waves indicates the velocity of light in free space must be much more than the phase velocity of these waves \cite{D.D.S, A.M}.
\section{Spectroscopy of magnetostatic modes of a rectangular waveguides filled with a medium of SRRs}
Now, let us consider a metallic rectangular waveguide of lateral dimensions $w$ and $h$ in the $y$ and $z$ directions, respectively, as shown in panel (b) of Fig. \ref{fig.1}. Initially assume that the waveguide is of infinite length and is filled with a medium of SRRs. It is our purpose to determine the various magnetostatic modes that can exist inside this rectangular magnetostatic waveguide. In this case, the wave equation for the magnetostatic potential inside the
waveguide can be written as  \b \label{10} \left[ \dfrac{\partial^{2}}{\partial x^{2}} +\mu_{yy}\dfrac{\partial^{2}}{\partial y^{2}}+\dfrac{\partial^{2}}{\partial z^{2}} \right]\psi=0 \;.\e
Again, the solution to Eq. \eqref{10} can be obtained by using the separation of variables method. In general, the solution to $\psi(x,y,z)$ for traveling magnetostatic waves
in the $+x$-direction can be written as 
\begin{multline} \label{11}
\psi(x,y,z)=\left( C \sin \kappa y+D \cos \kappa y\right)\\\times\left( E \sin \alpha z+F \cos \alpha z\right)\exp(-j\beta x)  \;.\end{multline}
Note that $C$, $D$, $E$, and $F$ are constants. Substituting this equation into Eq. \eqref{10}, we get \begin{figure}[!htb]
\centering
\includegraphics[width=9cm,clip]{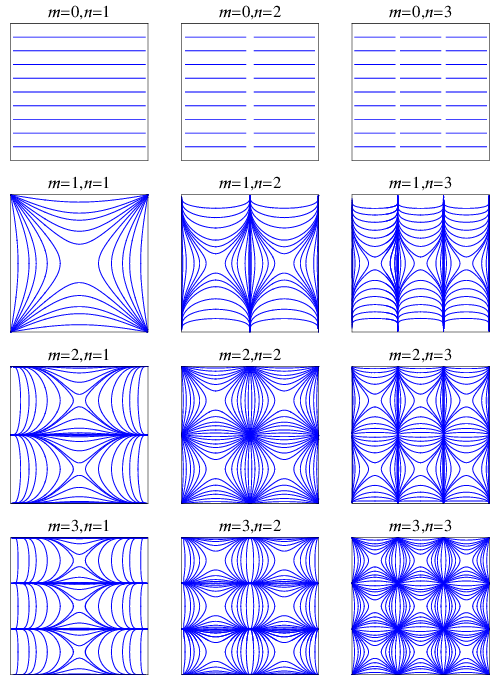}
\caption{Snapshot of the magnetic field patterns for the magnetostatic modes $(m=[0,1,2,3],n=[1,2,3])$ of a metallic square waveguide ($h=w$) filled with a medium of SRRs. }
\label{fig.6}
\end{figure} 
\b \label{12} \alpha^{2}+\beta^{2}+\mu_{yy}\kappa^{2}=0 \;.\e Here, the boundary conditions on the problem are that normal components of magnetic field vanish at the four metallic walls. Hence, we have  
\b \label{13}\dfrac{\partial \psi}{\partial y}\bigg\vert_{y=0,w}=0\;,\;\;\;\;\;\;\;\;\;\;\dfrac{\partial \psi}{\partial z}\bigg\vert_{z=0,h}=0\;. \e   To satisfy the boundary conditions for magnetostatic modes, we choose $ C=0=E$, $ \kappa=n \pi /w$, and $\alpha=m \pi /h$, where each integer $m$ and $n$, except $m=0=n$, specifies a mode. Hence, the magnetic
potential is
\b \label{15}
\psi(x,y,z)= A_{mn} \cos \left( \dfrac{n \pi y}{w}\right)  \cos \left( \dfrac{m \pi z}{h}\right)  \exp(-j\beta x) \;,\e
with $m=0,1,2,\cdots$; $n=0,1,2,\cdots$; $m=0=n$ excepted. In Eq. \eqref{15} the case $m=0=n$  is excluded because for that $\psi$  is a constant and all the components of magnetic field vanish; thus a trivial solution. The dispersion equation for the modes can be written as  
\b \label{16}  \beta^{2} = -\left(\dfrac{m\pi  }{h}\right)^{2}- \mu_{yy}\left(\dfrac{n\pi  }{w}\right) ^{2}\;.\e 
From the above dispersion relation it is easy to conclude that $n=0$ is not acceptable and should be excluded. Since there are infinite combinations of $m$ and $n$ ($n\neq 0$), an infinite number of magnetostatic modes can be found. Substituting Eq. \eqref{2} into Eq. \eqref{16}, we find  
\b \label{17} \omega=\omega_{0}\left(\dfrac{1+\left(\dfrac{n \pi }{\beta w} \right) ^{2}+\left(\dfrac{m \pi }{\beta h} \right) ^{2}}{1+(1-F)\left(\dfrac{n \pi }{\beta w} \right) ^{2}+\left(\dfrac{m \pi }{\beta h} \right) ^{2}}\right) ^{1/2}  \;.\e
 Also, for the cutoff frequency of a given $(m,n)$ mode, we find  
\b \label{18} \left( \omega_{c}\right) _{mn}=\omega_{0}\left( \dfrac{\left(\dfrac{n \pi }{w} \right) ^{2}+\left(\dfrac{m \pi }{h} \right) ^{2}}{(1-F)\left(\dfrac{n \pi }{w} \right) ^{2}+\left(\dfrac{m \pi }{h} \right) ^{2}} \right) ^{1/2} \;.\e
For $m=0$ and $n=1,2,3,\cdots$ we have $\left( \omega_{c}\right) _{0n}=\omega_{0}/\sqrt{1-F}$.

The dispersion curves for magnetostatic modes $(m,n)$ ($n=1,2,3$ and $m=0,1,2,3$) of a metallic square waveguide ($h = 2a = w$) filled with a medium of SRRs with $\omega_{0}=4$GHz and $F=0.56$ are depicted Fig. \ref{fig.5}. Note that by using the full set of Maxwell's equations, Eq. \eqref{16} for TE$_{mn}$ modes related to the $y$-axis should be read as (see Appendix)
 \b \label{5fff}  \beta^{2} = -\left(\dfrac{m\pi  }{h}\right)^{2}- \mu_{yy}\left[ \left(\dfrac{n\pi  }{w}\right) ^{2}-\dfrac{\omega^{2}}{c^{2}}\right] \;.\e
The mode patterns (field lines) are also of interest, where Fig. \ref{fig.6} shows sketches of cross-sectional mode patterns for some of the magnetostatic modes of a square waveguide, where results agree quite well with the results using the full set of Maxwell's equations for the large values of propagation constant (not shown here). In a rectangular waveguide, the field lines are bent from the location shown, but they retain the same general appearance. By comparing the results in figure 4 of \cite{C.L271} with Fig. \ref{fig.6} in the present work, one can conclude that the magnetic field pattern of magnetostatic mode $(m, n)$ is similar with magnetic field pattern of TE mode for a hollow waveguide.

Another part of the present analytical study is to find the power flow associated with the magnetostatic waves that may lead to further insight. To find the power that flows in the waveguide, the time-averaged of the power density directed along the axis of the waveguide is needed first. Then, we can obtain the power flow along the waveguide by integrating the power density in the axial direction over the cross-section of the waveguide. The power density along the $x$-direction that is delivered by a magnetostatic mode can be written as \cite{D.D.S, K.Y2170}
\b \label{19}
S_{x}=-\dfrac{\mu_{0}}{2}{\rm Re} \left[\psi\dfrac{\partial}{\partial t}\dfrac{\partial}{\partial x}\psi^{\ast}\right] \;, \e
in the complex-number representation, where $^{\ast}$ denotes complex conjugation, and ${\rm Re}$ denotes taking the real part. Use of \eqref{15} into \eqref{19}, the $x$-directed power density for the magnetostatic modes can be written as  
\b \label{20}
S_{x}=-\dfrac{\mu_{0}\omega k}{2} \vert A_{mn}\vert^2 \cos^{2} \left( \dfrac{n \pi y}{w}\right)  \cos^{2} \left( \dfrac{m \pi z}{h}\right) \;.\e
The power transmitted along the waveguide can be found by integrating Eq. \eqref{20} over the guide cross-section $\mathcal{A}=hw$. This gives \b \label{21}
P_{mn} =\int_{0}^{h}\int_{0}^{w} S_{x} dydz=-\dfrac{\mu_{0}\omega k}{8} hw \vert A_{mn}\vert^2\;.\e
Also, the time-averaged of energy density distribution associated with the waves can be written as \cite{K.Y2170}
\b \label{22}
u=\dfrac{\mu_{0}}{4} \nabla\psi^{\ast}\cdot\left[ \dfrac{d \left( \omega\underline{\mu}\right) }{d\omega}\cdot\nabla\psi\right] \;. \e
After substitution Eq. \eqref{15} into Eq. \eqref{22}, we obtain 
\begin{multline} \label{23}
u=\dfrac{\mu_{0}k^{2}}{4} \vert A_{mn}\vert^2 \cos^2 \left( \dfrac{n \pi y}{w}\right)  \cos^2 \left( \dfrac{m \pi z}{h}\right) \\+\dfrac{\mu_{0}}{4}\left( \dfrac{n \pi}{w}\right)^2\dfrac{d \left( \omega\mu_{yy}\right) }{d\omega} \vert A_{mn}\vert^2 \sin^2 \left( \dfrac{n \pi y}{w}\right)  \cos^2 \left( \dfrac{m \pi z}{h}\right)\\+\dfrac{\mu_{0}}{4}\left( \dfrac{m \pi}{h}\right)^2 \vert A_{mn}\vert^2 \cos^2 \left( \dfrac{n \pi y}{w}\right)  \sin^2 \left( \dfrac{m \pi z}{h}\right)  \;, \end{multline}
The associated storage energy is obtained by integrating Eq. \eqref{23} over the cross-section $\mathcal{A}$ of the waveguide, as
 \b \label{24}
U_{mn}=\dfrac{\mu_{0}}{16}wh\omega\vert A_{mn}\vert^2 \left( \dfrac{n \pi}{w}\right)^2\dfrac{d\mu_{yy} }{d\omega}      \;.\e
Note that using the simple formula $v_{\mathrm{g}}=d\omega/dk$ and Eq. \eqref{16}, the group velocity of magnetostatic modes of the system can be simply computed. However, in the absence of the damping effects, the group velocity of the waves is also equal with the ratio of the power flow 
to the storage energy, such as 
\b \label{25} v_{\mathrm{g}}=\dfrac{P_{mn} }{U_{mn} }=-2k   \left( \dfrac{n \pi}{w}\right)^{-2}\left( \dfrac{d\mu_{yy} }{d\omega}\right) ^{-1} \;,\e 
that is a manifestation of self-consistency and general validity of presented results in the magnetostatic theory.

\section{Conclusion}
In summary, we have investigated the existence of magnetostatic waves in a long, metallic rectangular waveguide filled with a metamaterial slab composed of SRRs as an anisotropic uniaxial $\mu$-negative
medium. We have obtained general expression for dispersion equation of the magneto-quasistatic TE waves and then presented the mode patterns in such magnetostatic waveguides. Physically, the present magnetostatic modes owe their existence to the anisotropic property of
the system, where in the absence of this property, these magnetostatic modes are disappeared. We have verified the obtained results by showing that group velocity of the waves is the same as energy velocity (i.e., the ratio of the power flow to the storage energy). Also, we have carried out rigorous electromagnetic analysis, which agree quite well with the magnetostatic results for the large propagation constant. Because of the possibility of magnetostatic waves propagation in a rectangular waveguides filled with  a metamaterial composed of SRRs, they may be used in the development of new waveguides using guided magnetostatic waves.

\section*{AUTHOR DECLARATIONS}
\subsection*{Conflict of Interest}

The authors have no conflicts to disclose.

\subsection*{Author Contributions}

\textbf{Afshin Moradi:} Project administration (lead); Conceptualization (lead); Investigation (lead);
Methodology (lead); Validation (lead); Formal analysis (lead); Software (lead); Writing - original draft
(lead); Writing – review and editing (equal). \textbf{Mohammed M. Bait-Suwailam:} Conceptualization (supporting); Validation (supporting); Formal analysis (supporting); Writing – review and editing (equal).

\subsection*{DATA AVAILABILITY} 

The data that supports the findings of this study are available within the article.

\renewcommand{\theequation}{A-\arabic{equation}} 
\setcounter{equation}{0}
\section*{Appendix: The rigorous electromagnetic analysis}
In the general case, let us consider a biaxial medium whose principal axes coincide with the axes of the Cartesian coordinate system $(x, y, z)$. The relative permittivity and permeability tensors $\underline{\varepsilon}$ and $\underline{\mu}$ can be represented by diagonal matrices
\begin{equation}\label{A1} \underline{\varepsilon}(\omega)=\left(%
\begin{array}{ccc}
\varepsilon_{xx} & 0 & 0\\
  0 & \varepsilon_{yy} & 0\\
  0 & 0 & \varepsilon_{zz}\\
\end{array}%
\right)\;,\end{equation}
and \begin{equation}\label{A2} \underline{\mu}(\omega)=\left(%
\begin{array}{ccc}
\mu_{xx} & 0 & 0\\
  0 & \mu_{yy} & 0\\
  0 & 0 & \mu_{zz}\\
\end{array}%
\right)\;.\end{equation}
Now, contrary to the assumption in \cite{S.H110, F.Y.M3400}, where the authors considered only the propagation of TE$_{m0}$ modes related to the waveguide axis (i.e., the $x$-axis in Fig. \ref{fig.1}), let us suppose that the axis of decomposition of TM and TE modes is along the $y$-axis. 

For the benefit of the reader of the present work, let us note that Meng \textit{et al.} \cite{F.Y.M3400} analyzed rectangular waveguides loaded with anisotropic metamaterials to assess the controllability of transmission characteristics of the involved electromagnetic waves. In this way, by assuming the existence of TE waves related to the waveguide axis in the mentioned system, at first, they considered the propagation of TE$_{mn}$ modes (see Eqs. (4), (6), (7), (8), and (9) in \cite{F.Y.M3400}) and in the following, they investigated the propagation of TE$_{m0}$ modes. However, in the general case, hybrid wave propagation should be expected for the rectangular waveguides loaded with anisotropic metamaterials, and therefore TE$_{mn}$ waves are unable to propagate. Therefore, Eqs. (4), (6), (7), (8), and (9) in \cite{F.Y.M3400} are incorrect (see \cite{A.MTAP}). 

Substituting electric and magnetic fields $\textbf{E}$ and $\textbf{H}$ describing a wave traveling in the $x$-direction
\begin{equation}\label{A3} \textbf{E}(x,y,x)=\left(%
\begin{array}{ccc}
E_{x}(y,z) \\
 E_{y}(y,z) \\
 E_{z}(y,z)\\
\end{array}%
\right)e^{j(\omega t-\beta x)}\;,\end{equation}
\begin{equation}\label{A4} \textbf{H}(x,y,x)=\left(%
\begin{array}{ccc}
H_{x}(y,z) \\
H_{y}(y,z) \\
H_{z}(y,z)\\
\end{array}%
\right)e^{j(\omega t-\beta x)}\;,\end{equation}
into Maxwell's equations 
\b \label{A5}\nabla\times\textbf{E}=-j\omega\mu_{0}\underline{\mu}\textbf{H} \;, \e
\b \label{A6}\nabla\times\textbf{H}=j\omega\varepsilon_{0}\underline{\varepsilon}\textbf{E} \;, \e
we obtain
\b \label{A7}\dfrac{\partial E_{z}}{\partial y}-\dfrac{\partial E_{y}}{\partial z}=-j\omega\mu_{0}\mu_{xx}H_{x} \;, \e
\b \label{A8}j\beta E_{z}+\dfrac{\partial E_{x}}{\partial z}=-j\omega\mu_{0}\mu_{yy}H_{y} \;, \e
\b \label{A9}j\beta E_{y}+\dfrac{\partial E_{x}}{\partial y}=j\omega\mu_{0}\mu_{zz}H_{z} \;, \e
\b \label{A10}\dfrac{\partial H_{z}}{\partial y}-\dfrac{\partial H_{y}}{\partial z}=j\omega\varepsilon_{0}\varepsilon_{xx}E_{x} \;, \e
\b \label{A11}j\beta H_{z}+\dfrac{\partial H_{x}}{\partial z}=j\omega\varepsilon_{0}\varepsilon_{yy}E_{y} \;, \e
\b \label{A12}j\beta H_{y}+\dfrac{\partial H_{x}}{\partial y}=-j\omega\varepsilon_{0}\varepsilon_{zz}E_{z} \;. \e
Thus, following the method of determinants, we obtain $E_{x}$, $E_{z}$, $H_{x}$, and $H_{z}$ with respect to components $E_{y}$ and $H_{y}$, as
\b \label{A13}E_{x}=\dfrac{j}{\kappa^{2}-\varepsilon_{xx}\mu_{zz}\dfrac{\omega^{2}}{c^{2}}}\left( \beta\dfrac{\partial E_{y}}{\partial y}-\omega\mu_{0}\mu_{zz}\dfrac{\partial H_{y}}{\partial z}\right)  \;, \e
\b \label{A14}H_{x}=\dfrac{j}{\kappa^{2}-\varepsilon_{zz}\mu_{xx}\dfrac{\omega^{2}}{c^{2}}}\left( \beta\dfrac{\partial H_{y}}{\partial y}+\omega\varepsilon_{0}\varepsilon_{zz}\dfrac{\partial E_{y}}{\partial z}\right) \;, \e
\b \label{A15}E_{z}=\dfrac{1}{\kappa^{2}-\varepsilon_{zz}\mu_{xx}\dfrac{\omega^{2}}{c^{2}}}\left( \omega\beta\mu_{0}\mu_{xx} H_{y}-\dfrac{\partial^{2}E_{y}}{\partial y \partial z}\right)  \;, \e
\b \label{A16}H_{z}=\dfrac{-1}{\kappa^{2}-\varepsilon_{xx}\mu_{zz}\dfrac{\omega^{2}}{c^{2}}}\left( \omega\beta\varepsilon_{0}\varepsilon_{xx} E_{y}+\dfrac{\partial^{2}H_{y}}{\partial y \partial z}\right)  \;, \e
where we have used $\partial^{2}/\partial y^{2}=-\kappa^{2}$. Substituting Eqs. \eqref{A13} and \eqref{A15} into Eq. \eqref{A8}, and Eqs. \eqref{A14} and \eqref{A16} into Eq. \eqref{A11} we get
\begin{multline} \label{A17}\dfrac{\mu_{zz}}{\kappa^{2}-\varepsilon_{xx}\mu_{zz}\dfrac{\omega^{2}}{c^{2}}}\dfrac{\partial^{2}H_{y}}{\partial z^{2}}-\left(\mu_{yy}+\dfrac{\mu_{xx}\beta^{2}}{\kappa^{2}-\varepsilon_{zz}\mu_{xx}\dfrac{\omega^{2}}{c^{2}}} \right)H_{y}=\\ \dfrac{\omega\beta\varepsilon_{0}\left(\varepsilon_{xx}\mu_{zz}-\varepsilon_{zz}\mu_{xx}\right) }{\left( \kappa^{2}-\varepsilon_{xx}\mu_{zz}\dfrac{\omega^{2}}{c^{2}}\right)\left( \kappa^{2}-\varepsilon_{zz}\mu_{xx}\dfrac{\omega^{2}}{c^{2}}\right) }\dfrac{\partial^{2}E_{y}}{\partial y \partial z} \;, \end{multline}
\begin{multline} \label{A18}\dfrac{\varepsilon_{zz}}{\kappa^{2}-\varepsilon_{zz}\mu_{xx}\dfrac{\omega^{2}}{c^{2}}}\dfrac{\partial^{2}E_{y}}{\partial z^{2}}-\left(\varepsilon_{yy}+\dfrac{\varepsilon_{xx}\beta^{2}}{\kappa^{2}-\varepsilon_{xx}\mu_{zz}\dfrac{\omega^{2}}{c^{2}}} \right)E_{y}=\\ \dfrac{\omega\beta\mu_{0}\left( \varepsilon_{xx}\mu_{zz}-\varepsilon_{zz}\mu_{xx}\right) }{\left( \kappa^{2}-\varepsilon_{xx}\mu_{zz}\dfrac{\omega^{2}}{c^{2}}\right)\left( \kappa^{2}-\varepsilon_{zz}\mu_{xx}\dfrac{\omega^{2}}{c^{2}}\right) }\dfrac{\partial^{2}H_{y}}{\partial y \partial z} \;. \end{multline}
In general, Eqs. \eqref{A17} and \eqref{A18} may be satisfied by the same value of $\beta$ if $E_{y}\neq 0$ and $H_{y}\neq 0$. Hence, the so-called anomalous modes having all the six field components would be possible. However, let us explore the possibility of TE$_{y}$ and TM$_{y}$ wave propagation. If we assume $E_{y}= 0$ (TE$_{y}$ modes), then Eqs. \eqref{A17} and \eqref{A18}, reduce to 
\b \label{A19}\dfrac{\mu_{zz}}{\kappa^{2}-\varepsilon_{xx}\mu_{zz}\dfrac{\omega^{2}}{c^{2}}}\dfrac{\partial^{2}H_{y}}{\partial z^{2}}-\left(\mu_{yy}+\dfrac{\mu_{xx}\beta^{2}}{\kappa^{2}-\varepsilon_{zz}\mu_{xx}\dfrac{\omega^{2}}{c^{2}}} \right)H_{y}=0 \;, \e
\b \label{A20} \dfrac{\omega\beta\mu_{0}\left( \varepsilon_{xx}\mu_{zz}-\varepsilon_{zz}\mu_{xx}\right) }{\left( \kappa^{2}-\varepsilon_{xx}\mu_{zz}\dfrac{\omega^{2}}{c^{2}}\right)\left( \kappa^{2}-\varepsilon_{zz}\mu_{xx}\dfrac{\omega^{2}}{c^{2}}\right) }\dfrac{\partial^{2}H_{y}}{\partial y \partial z}=0 \;. \e
Equations \eqref{A19} and \eqref{A20} cannot be simultaneously satisfied by the same value of $\beta$ unless 
in two special cases. In the first case, decoupling takes place if  
\b \label{A21} \dfrac{\partial H_{y}}{\partial y}=0\;,\;\;\;\mathrm{or}\;\;\; \dfrac{\partial H_{y}}{\partial z}=0\;.\e
In the second case, decoupling takes place if material parameters satisfy the condition
\b \label{A22} \varepsilon_{xx}\mu_{zz}=\varepsilon_{zz}\mu_{xx}\;,\e
that is simply possible for our present system, because we have $\varepsilon_{xx}=\varepsilon_{zz}=1$ and $\mu_{xx}=\mu_{zz}=1$. Therefore, we obtain
\b \label{A23}\dfrac{\partial^{2}H_{y}}{\partial z^{2}}-\left(\beta^{2}+\mu_{yy}\left[ \kappa^{2}-\dfrac{\omega^{2}}{c^{2}}\right]  \right)H_{y}=0 \;. \e
If we set $\partial/\partial z=0$, we get Eq. \eqref{5f}, and by using $\partial^{2}/\partial z^{2}=-\alpha^{2}$, Eq. \eqref{5fff} can be obtained.

Similarly, if we assume that (TM$_{y}$ modes), then Eqs. \eqref{A17} and \eqref{A18}, reduce to 
\b \label{A24} \dfrac{\omega\beta\varepsilon_{0}\left(\varepsilon_{xx}\mu_{zz}-\varepsilon_{zz}\mu_{xx}\right) }{\left( \kappa^{2}-\varepsilon_{xx}\mu_{zz}\dfrac{\omega^{2}}{c^{2}}\right)\left( \kappa^{2}-\varepsilon_{zz}\mu_{xx}\dfrac{\omega^{2}}{c^{2}}\right) }\dfrac{\partial^{2}E_{y}}{\partial y \partial z}=0 \;, \e
\b \label{A25}\dfrac{\varepsilon_{zz}}{\kappa^{2}-\varepsilon_{zz}\mu_{xx}\dfrac{\omega^{2}}{c^{2}}}\dfrac{\partial^{2}E_{y}}{\partial z^{2}}-\left(\varepsilon_{yy}+\dfrac{\varepsilon_{xx}\beta^{2}}{\kappa^{2}-\varepsilon_{xx}\mu_{zz}\dfrac{\omega^{2}}{c^{2}}} \right)E_{y}=0 \;. \e
Again, Eqs. \eqref{A24} and \eqref{A25} cannot be simultaneously satisfied unless material parameters satisfy the condition shown by Eq. \eqref{A22}, or if
\b \label{A26} \dfrac{\partial E_{y}}{\partial y}=0\;,\;\;\;\mathrm{or}\;\;\; \dfrac{\partial E_{y}}{\partial z}=0\;.\e
For our present system, we have $\varepsilon_{xx}=\varepsilon_{yy}=\varepsilon_{zz}=1$ and $\mu_{xx}=\mu_{zz}=1$, then we obtain
 \b \label{A27}\dfrac{\partial^{2}E_{y}}{\partial z^{2}}-\left(\kappa^{2}+\beta^{2}-\dfrac{\omega^{2}}{c^{2}} \right)E_{y}=0 \;. \e
This equation describes the propagation of TM$_{y}$ waves in hollow waveguides and is well-known in many electromagnetic textbooks. In this case, the system cannot support quasi-static waves.

\end{document}